\documentclass[11pt]{article}
\begin{document}
\setcounter{page}{1}
\newcommand{\FTS}[2]{\frac{{\textstyle #1}}{{\textstyle #2}}}
\def\NN{\hbox{\bf N}}
\def\RR{\hbox{\bf R}}
\def\CC{\hbox{\bf C}}
\def\ZZ{\hbox{\bf Z}}
\def\QQ{\hbox{\bf Q}}
\def\theorem{{\sc Theorem}\ }
\def\remark{{\sc Remark}\ }
\def\definition{{\sc Definition}\ } 
%%%%%%%%%%%%%%%%%%%%%%
\def\pni{\par \noindent}
\def\vsh{\smallskip}
\def\vs{\medskip}
\def\vvs{\bigskip}
\def\vvvs{\bigskip\medskip} %% {\vskip 1.5truecm}
\def\vsp{\par\noindent}
\def\vsn{\vsh\pni}
\def\cen{\centerline}
\def\ra{\item{a)\ }} \def\rb{\item{b)\ }}   \def\rc{\item{c)\ }}
%%%%%%%%%%%%%%%%%%%%%%%%%%%%%%%%%%
%% \begin{center}
\cen{FRACALMO PRE-PRINT: \   {\bf www.fracalmo.org}}
\vsh
\cen{{\bf Fractional Calculus and Applied Analysis,
  Vol. 2 No 4 (1999), pp. 383-414}}
\vsh
\cen{An International Journal for Theory and Applications \ ISSN 1311-0454}
\vsh
\cen{{\bf www.diogenes.bg/fcaa/}}
\vs
\hrule
% \end{center}
%%%%%%%%%%%%%%%%%%%%%%%%%%%%%%%%%%%%%%%%%%%%%%%%%%%%%%%%%%%%%%%%%%%%%%%%%
%%BEGINNING OF TEXT
%%%%%%%%%%%%%%%%%%%%%%%%%%%%%%%%%%%%%%%%%%%%%%%%%%%%%%%%%%%%%%%%%%%%%%%%%
   \vskip 0.50truecm

\font\title=cmbx12 scaled\magstep2
\font\bfs=cmbx12 scaled\magstep1

\begin{center}

{\title Analytical properties and applications}
\vs

{\title of the Wright function}

\vvs

 {Rudolf GORENFLO}$^{(1)}$,
{Yuri LUCHKO}$^{(2)}$,
{Francesco MAINARDI} $^{(3)}$

%%%%%%%%%%%%%%%%

\vs

$\null^{(1)}$
 First Mathematical Institute,
 %%%Department of Mathematics and  Informatics,
 Free   University of Berlin, \\
 Arnimallee  3, D-14195 Berlin, Germany \\
%% Tel: +49-30-838.75.??? $\;$ Fax: +49-30-838.75.??? $\;$ \\
E-mail: {\tt gorenflo@mi.fu-berlin.de}
\\ [0.25 truecm]

$\null^{(2)}$
 Department of Mathematics (II),\\
Technical University of Applied Sciences Berlin, \\
Luxemburger Str. 10, D-13353 Berlin, Germany \\
%% Phone: +49 30 4504 2247 \\
E-mail: {\tt luchko@tfh-berlin.de}
\\ [0.25 truecm]

$\null^{(3)}$
 Department of Physics, University of Bologna, and INFN, \\
              %%%  Sezione di Bologna, \\
Via Irnerio 46, I-40126 Bologna, Italy \\
%% Tel: +39-051.2091098 $\;$ Fax: +39-051.247244 $\;$\\
E-mail: {\tt francesco.mainardi@unibo.it} 
\\ [0.25 truecm]

\subsection*{Abstract}
\end{center}

\noindent
The entire function (of $z$)
%\begin{equation}
%\label{1}
\[
\phi(\rho,\beta;z)=\sum_{k=0}^\infty {z^k\over k! \Gamma(\rho k+\beta)},\
\rho >-1,\ \beta\in \CC,
\]
%\end{equation}
named after the British mathematician E.M. Wright,
has appeared for the first time in the case $\rho >0 $
 in
connection with his investigations in the asymptotic theory of partitions.
Later on, it has found many other applications, first of all, in the
Mikusi\'nski operational calculus and in the theory of integral transforms
of Hankel type. Recently this
function has appeared in  papers related to  partial differential equations
of fractional order. Considering the boundary-value problems for the
fractional diffusion-wave equation, i.e., the linear  partial
integro-differential equation obtained from the classical diffusion or wave
equation by replacing the first- or second-order time derivative by a
fractional derivative of order $\alpha$ with $0<\alpha\le 2$, it was found
that the corresponding Green functions can be represented in terms of the
Wright function. Furthermore, extending the methods of Lie groups in
partial differential equations to the partial differential equations of
fractional order it was shown that some
of the group-invariant solutions of these equations  can be given in
terms of the Wright and the generalized Wright functions. In this
{\it survey paper\/} we consider some of the above mentioned applications
of the Wright function with special emphasis of its key role in
the partial differential equations of fractional order.

We also give some analytical tools for working with this function. Beginning
with the classical results of Wright about the asymptotics of this function,
we present other properties, including its
representations in terms of the special functions of the hypergeometric
type and the Laplace
transform pairs related to the Wright function. Finally, we
discuss recent results about distribution of zeros of the Wright
function, its order, type and indicator function, showing that this function
is an entire function of completely regular growth for every $\rho >-1$.

\vskip 6pt
\noindent
{\it Mathematics Subject Classification}: primary 33E20,
secondary 33C20, 30C15,
30D15, 26A33, 45J05, 45K05

\vskip 6pt
\noindent
{\it Key Words and Phrases}: Wright function, indicator function,
asymptotics of zeros, generalized hypergeometric function,
diffusion-wave equation, Green function,
scale-invariant solutions, Erd\'elyi-Kober
operators

\vskip 12pt

\section*{1. Introduction}

The purpose of this survey paper is to outline the fundamental role of
the Wright function in partial differential equations of fractional
order, to consider some other applications of this function and to give
its analytical properties including asymptotics and distribution of
its zeros. Partial differential equations of fractional order (FPDE)
are obtained by
replacing some (or all) derivatives in partial differential equations by
derivatives of fractional order (in Caputo, Riemann-Liouville or inverse
Riesz potential sense). Mathematical aspects of the boundary-value problems
for some FPDE
have been treated in papers by several authors including Engler
\cite{Eng1}, Fujita \cite{Fu1}, Gorenflo and Mainardi \cite{Go1}, Mainardi
\cite{Ma1}-\cite{Ma3}, Podlubny \cite{Pod}, Pr\"uss \cite{Pr1}, Saichev and
Zaslavsky \cite{Sai1}, Samko et al. \cite{Samko}, Schneider and Wyss
\cite{Sc1} and by Wyss \cite{Wy1}.

From the other side, some FPDE were
successfully used for modelling relevant physical processes (see,
for example  Giona and Roman \cite{GionaRoman92}, Hilfer \cite{Hilfer95},
Mainardi \cite{Mai}, Metzler et al. \cite{Metzler94}, Nigmatullin
\cite{Nig}, Pipkin \cite{Pip}, Podlubny \cite{Pod} and references there).
In applications, special types of solutions, which
are invariant under some subgroup of the full symmetry group of the given
equation (or for a system of equations) are especially important.

Recently, the scale-invariant solutions for time-fractional
diffusion-wave equation (with the fractional derivative in the
Riemann-Liouville sense) and for the more general time- and space-fractional
partial differential equation (with the Riemann-Liouville space-fractional
derivative of  order $\beta\le 2$ instead of the second order space
derivative)  have been presented by Buckwar and
Luchko \cite{BucLuc} and by Luchko and Gorenflo \cite{Luc}, respectively.
The case of the time-fractional diffusion-wave equation with the Caputo
fractional derivative has been considered by Gorenflo, Luchko and Mainardi
\cite{GorLucMai}.
%%%%%%%%%%%%%%%%%%%%%%%%%%%

The plan of the paper is as follows. In Section 2, following the
papers by Djrbashian and Bagian \cite{DjrBag}, Gaji\'c and Stankovi\'c
\cite{Gaj},  Luchko and Gorenflo \cite{Luc}, Mainardi
\cite{Ma3}, Mainardi and Tomirotti \cite{MaiTom}, Mikusi\'nski \cite{Mik},
Pathak \cite{Pat}, Pollard \cite{Pol}, Stankovi\'c \cite{Sta}, and Wright
\cite{Wr1}, \cite{Wright}, we recall the main properties of the Wright
function including its integral representations, asymptotics,
representations in terms of the special functions of the hypergeometric
type and the Laplace
transform pairs related to the Wright function. Finally, we
discuss new results about distribution of zeros of the Wright
function, its order, type and indicator function, showing that this function
is an entire function of completely regular growth for every $\rho >-1$.

In Section 3, we outline some applications of the Wright function,
beginning with the results by Wright \cite{Wri1} in the asymptotic theory
of partitions. Special attention is given to the key role of the
Wright function in the theory of FPDE. Following Gorenflo,
Mainardi and Srivastava \cite{GorMaiSri} and
Mainardi \cite{Ma1}-\cite{Ma3}, we consider in details the
boundary-value problems of Cauchy and
signalling type for the fractional diffusion-wave equation,  showing
that the corresponding Green functions can be represented in terms of the
Wright function. We present also some results from  Buckwar and Luchko
\cite{BucLuc}, Gorenflo, Luchko
and Mainardi \cite{GorLucMai}, Luchko and Gorenflo \cite{Luc}
concerning the extension of  the methods of the Lie groups in partial
differential equations to FPDE.
It will be shown that some of the group-invariant solutions of FPDE can
be given in terms of the Wright and the generalized Wright functions.

We remark finally that the present review is essentially based on our
original works. For the other
applications of the Wright function, including Mikusi\'nski's operational
calculus and the theory of integral transforms of Hankel type we refer, for
example,
to Kiryakova \cite{Kir}, Kr\"atzel \cite{Kra}, Mikusi\'nski \cite{Mik},
and Stankovi\'c \cite{Sta}.

\section*{2. Analytical properties}

\subsection*{2.1. Asymptotics}

Probably the most important characteristic of a special function
is its asymptotics. In the case of an entire function there are deep
relations between its asymptotic behaviour in the neighbourhood
of its only singular point -- the essential singularity at $z=\infty$ --
and other properties of this function, including distribution of its zeros
(see, for example, Evgrafov \cite{Evg}, Levin \cite{Lev}).
It follows from the Stirling asymptotic formula for the
gamma function that the Wright function
\begin{equation}
\label{Wright}
\phi(\rho,\beta;z)=\sum_{k=0}^\infty {z^k\over k! \Gamma(\rho k+\beta)},\
\rho >-1,\ \beta\in \CC,
\end{equation}
is an entire function of $z$ for $\rho >-1$ and, consequently, as we will see
in the later parts of our survey, some elements of the general
theory of entire functions can be applied.

The complete picture of the asymptotic behaviour of the Wright function for
large values of $z$  was given by
Wright \cite{Wr1} in the case $\rho > 0$ and by Wright \cite{Wright} in the
case $-1<\rho<0$. In both cases
he used the method of steepest descent and the integral representation
\begin{equation}
\label{repwr}
\phi(\rho,\beta;z)={1\over 2\pi i}\int_{{\mathrm Ha}}
e^{\zeta+z \zeta^{-\rho}} \zeta^{-\beta}\, d\zeta,\ \rho >-1,\ \beta \in \CC
\end{equation}
where ${\mathrm Ha}$ denotes the Hankel path in the $\zeta$-plane with a
cut along the negative real semi-axis $\arg \zeta =\pi$. Formula
(\ref{repwr}) is obtained by substituting the
Hankel representation for the reciprocal of the gamma function
\begin{equation}
\label{gamma}
{1\over\Gamma(s)}={1\over 2\pi i}
\int_{{\mathrm Ha}}
e^\zeta \zeta^{-s}\, d\zeta,\ s\in\CC
\end{equation}
for $s=\rho k +\beta$ into (\ref{Wright}) and changing the order of
integration and summation.

\bigskip

Let us consider at first the case $\rho >0$.

\theorem{2.1.1}
If $\rho >0,\ \arg (-z) = \xi,\ |\xi|\le \pi$,
and
\[
Z_1=(\rho|z|)^{1/(\rho +1)}e^{i(\xi+\pi)/(\rho+1)},\
Z_2=(\rho|z|)^{1/(\rho +1)}e^{i(\xi-\pi)/(\rho+1)},
\]
then we have
\begin{equation}
\label{a0_1}
\phi(\rho,\beta;z) = H(Z_1)+H(Z_2),
\end{equation}
where
$H(Z)$ is given by
\begin{equation}
\label{H}
H(Z) = Z^{{1\over 2}-\beta} e^{{1+\rho\over \rho}Z}\left\{ \sum_{m=0}^M
{(-1)^m a_m \over Z^m} +{\rm O}\left( {1\over |Z|^{M+1}}\right)\right\},\
Z\to \infty
\end{equation}
and the $a_m,\ m=0,1,\dots$, are defined as the coefficients of $v^{2m}$ in the
expansion of
\[
{\Gamma(m+{1\over 2})\over 2\pi} \left({2\over \rho +1}\right)^{m+{1\over 2
}}(1-v)^{-\beta}\{g(v)\}^{-2m-1}
\]
with
\[
g(v)=\left\{ 1+{\rho +2\over 3}v +{(\rho +2)(\rho +3)\over 3\cdot 4}v^2
+\dots\right\}^{{1\over 2}}.
\]

\bigskip

In particular, if $\beta\in \RR$ we get the asymptotic expansion
of the Wright function $\phi(\rho,\beta;-x)$ for $x\to +\infty$ in the form
\begin{equation}
\label{a0_2}
\phi(\rho,\beta;-x) = x^{p({1\over
2}-\beta)}e^{\sigma x^p \cos \pi p}
\cos\left( \pi p({1\over 2}-\beta) +
\sigma x^p \sin\pi p\right)\left\{ c_1+ O(x^{-p})\right\},
\end{equation}
where $p={1\over 1+\rho}$, $\sigma=(1+\rho)\rho^{-{\rho\over 1+\rho}}$
 and the constant $c_1$
can be exactly evaluated.

If we exclude from the consideration an arbitrary small angle containing
the negative real semi-axis, we get a simpler result.

\theorem{2.1.2}
If $\rho >0,\ \arg z = \theta,\ |\theta|\le \pi-\epsilon,\ \epsilon >0$, and
\[
Z=(\rho|z|)^{1/(\rho +1)}e^{i\theta/(\rho+1)},
\]
then we have
\begin{equation}
\label{a0}
\phi(\rho,\beta;z) = H(Z),
\end{equation}
where $H(z)$ is given by {\rm (\ref{H})}.

\bigskip

In the case $\rho =0$ the Wright function is reduced to the exponential
function with the constant factor $1/\Gamma(\beta)$:
\begin{equation}
\label{exp}
\phi(0,\beta;z)=\exp(z)/\Gamma(\beta),
\end{equation}
which turns out to vanish identically for $\beta =-n,\
n=0,1,\dots$.

To formulate the results for the case $-1<\rho <0$ we introduce some
notations. Let
\begin{equation}
\label{cond1}
y=-z,\
-\pi<\arg
z\le\pi,\ -\pi<\arg y\le \pi,
\end{equation}
and let
\begin{equation}
\label{cond2}
Y=(1+\rho) \left((-\rho)^{-\rho} y\right)^{1/(1+\rho)}.
\end{equation}

\theorem{2.1.3}
If $-1<\rho<0,\ |\arg y| \le \min\{ {3\over 2}\pi(1+\rho),\pi\}
-\epsilon,\ \epsilon >0$, then
\begin{equation}
\label{a1}
\phi(\rho,\beta;z) = I(Y),
\end{equation}
where
\begin{equation}
\label{I}
I(Y)=Y^{{1\over 2}-\beta}e^{-Y}\left\{ \sum_{m=0}^{M-1} A_m Y^{-m} +
{\rm O}(Y^{-M})\right\}, \ Y\to \infty,
\end{equation}
and the coefficients $A_m,\ m=0,1\dots$ are defined by the asymptotic
expansion
\begin{eqnarray}
{\Gamma(1-\beta-\rho t)\over 2\pi (-\rho)^{-\rho t}(1+\rho)^{(1+\rho)
(t+1)}\Gamma(t+1)} & = & \sum_{m=0}^{M-1} {(-1)^m A_m\over \Gamma((1+\rho)t +
\beta +{1\over 2}+m)} \nonumber \\
 & & +{\rm O}\left( {1\over \Gamma((1+\rho)t +
\beta +{1\over 2}+M)}\right), \nonumber
\end{eqnarray}
valid for  $\arg t,\ \arg(-\rho t),$ and $\arg(1-\beta-\rho t)$ all lying
between $-\pi$ and $\pi$ and $t$ tending to infinity.

\bigskip

If $-1/3\le\rho<0$, the only region not covered by Theorem 2.1.3 is the
neighbourhood of the positive real semi-axis. Here we have the following
result.

\theorem{2.1.4}
If $-1/3<\rho<0,\ |\arg z| \le \pi(1+\rho)
-\epsilon,\ \epsilon >0$, then
\begin{equation}
\label{a1_1}
\phi(\rho,\beta;z) = I(Y_1)+I(Y_2),
\end{equation}
where
$I(Y)$ is defined by {\rm (\ref{I})},
\begin{equation}
\label{Y12}
Y_1=(1+\rho) \left((-\rho)^{-\rho} z e^{\pi i}\right)^{1/(1+\rho)},
Y_2=(1+\rho) \left((-\rho)^{-\rho} z e^{-\pi i}\right)^{1/(1+\rho)},
\end{equation}
hence
\[
Y_1=Y\ {\rm if}\ -\pi<\arg z\le 0,\ \ {\rm and}\ \ Y_2=Y\ {\rm if}\ 0<\arg z\le \pi.
\]

\bigskip

As a consequence we get the asymptotic expansion
of the Wright function $\phi(\rho,\beta;x)$ for $x\to +\infty$ in the case
$-1/3<\rho<0, \ \beta\in \RR$ in the form:
\begin{equation}
\label{a1_2}
\phi(\rho,\beta;x) = x^{p({1\over 2}-\beta)} e^{-\sigma x^p\cos
\pi p}\cos (\pi p({1\over 2}-\beta) -\sigma x^p\sin \pi
p)\left\{c_2 +O(x^{-p})\right\},
\end{equation}
where $p={1\over 1+\rho}$, $\sigma=(1+\rho)(-\rho)^{-{\rho\over 1+\rho}}$
 and the constant $c_2$
can be exactly evaluated.
When $-1<\rho <-1/3$,
there is a region of the plane in which the expansion is algebraic.

\theorem{2.1.5}
If $-1<\rho<-1/3,\ |\arg z| \le  {1\over 2}\pi(-1-3\rho)
-\epsilon,\ \epsilon >0$, then
\begin{equation}
\label{a2}
\phi(\rho,\beta;z) = J(z),\ z\to \infty,
\end{equation}
where
\begin{equation}
\label{J}
J(z)=\sum_{m=0}^{M-1} {z^{(\beta-1-m)/(-\rho)}\over (-\rho)\Gamma(m+1)
\Gamma(1+(\beta-m-1)/(-\rho))} +{\rm O}(z^{{\beta-1-M\over -\rho}}).
\end{equation}

\bigskip

Finally, the asymptotic expansions of the Wright function
in the
neighbourhood of the positive real semi-axis in the case $\rho =-1/3$
and in the neighbourhood of the lines
$\arg z=\pm {1\over 2}\pi(-1-3\rho)$ when $-1<\rho <-1/3$ are given by the
following
results by Wright.

\theorem{2.1.6}
If $\rho=-1/3,\ |\arg z| \le \pi(1+\rho)
-\epsilon,\ \epsilon >0$, then
\begin{equation}
\label{a1_3}
\phi(\rho,\beta;z) = I(Y_1)+I(Y_2)+J(z),
\end{equation}
where
$I(Y)$ is defined by {\rm (\ref{I})}, $Y_1, Y_2$ by {\rm (\ref{Y12})}, and
$J(z)$ by {\rm (\ref{J})}.

\theorem{2.1.7}
If $-1<\rho<-1/3,\ |\arg z \pm {1\over 2}\pi(-1-3\rho)| \le  \pi
(1+\rho)-\epsilon,\ \epsilon >0$, then
\begin{equation}
\label{a2_1}
\phi(\rho,\beta;z) = I(Y)+J(z),
\end{equation}
where $I(Y)$ is defined by {\rm (\ref{I})} and $J(z)$ by
{\rm (\ref{J})}.

\bigskip

The results given above contain the complete description of the
asymptotic behaviour of the Wright function for
large values of $z$ and for all values of the parameters $\rho >-1,\
\beta\in \CC$.
We will use  them repeatedly in our further discussions.

\subsection*{2.2. Representations through hypergeometric functions}

Due to the relation
\begin{equation}
\label{Bess}
\phi(1,\nu+1;-{1\over 4}z^2)=\left({z\over 2}\right)^{-\nu}J_\nu(z).
\end{equation}
Wright considered the function $\phi(\rho,\beta;z)$ as a generalization
of the Bessel function $J_\nu(z)$.
In the general case of arbitrary  real $\rho>-1$ the Wright function
is a
particular case of the Fox $H$-function (\cite{GorLucMai},
\cite{GorMaiSri}, \cite[App. E]{Kir},  \cite[Chapter 1]{Sri}):
\begin{equation}
\label{Fox}
\phi(\rho,\beta;z)=H^{1,0}_{0,2}\left[ -z\left|
\begin{array}{c}
- \\
(0,1),(1-\mu,\rho)
\end{array}
\right. \right].
\end{equation}
Unfortunately, since the Fox H-function is a very general object this
representation is not especially informative. It turns out that
if $\rho$ is a positive rational number the Wright function can be
represented in terms
of the more familiar generalized hypergeometric functions.
Let be $\rho =n/m$ with positive integers $n$ and $m$.
Substituting $s=ms_1$ into (\ref{Fox}) and making use of the
Gauss-Legendre formula for the gamma function
\[
\Gamma(n z)=n^{n z -{1\over 2}}(2\pi)^{1-n\over
2}\prod_{k=0}^{n-1}\Gamma(z+{k\over n}),\ \ n=2,3,\dots\,
\]
we arrive at the representation
\begin{equation}
\phi({n\over m},\beta;z)=(2\pi)^{{n-m\over 2}} m^{{1\over 2}}n^{-\beta+{1\over 2}}
{1\over 2\pi i}\int\limits_{{\mathrm L}_{-\infty}}
{\prod\limits_{k=0}^{m-1}\Gamma(s_1 +{k\over m})\over
\prod\limits_{l=0}^{n-1}
\Gamma({\beta\over n}-s_1+{l\over n})}
\left({(-z)^m\over m^mn^n}\right)^{-s_1}ds_1,
\end{equation}
which is equivalent to the representation given by Pathak \cite{Pat} in
terms of the Meijer $G$-function (\cite[App. A]{Kir}, \cite[Chapter
4]{Marichev}). Here ${\mathrm L}_{-\infty}$ is a loop beginning and ending
at $-\infty$, encircling in the positive direction all the poles of
$\Gamma(s_1 +{k\over m})$, $k=0,\dots,m-1$, i.e., the points
$-{k\over m},\ -1-{k\over m},\dots$.
The residue theorem
and the relation (\cite[Chapter 3]{Marichev})
\[
{\mathrm {res}}_{z=-k} \Gamma(z) = {(-1)^k\over k!},\ k=0,1,2,\dots
\]
allow us to represent this integral as a sum of $m$ series of
hypergeometric type:
\[
\phi({n\over m},\beta;z)=(2\pi)^{{n-m\over 2}} {m^{{1\over 2}}\over
n^{\beta-{1\over 2}}}\sum_{p=0}^{m-1}\sum_{q=0}^\infty {(-1)^q\over q!}
{\prod\limits_{{k=0\atop k\not = p}}^{m-1}\Gamma(-q-{p-k\over m})
\over \prod\limits_{l=0}^{n-1}\Gamma({\beta\over n}+{p\over m}+q +{l\over
n})}\left({(-z)^m\over m^mn^n}\right)^{q+{p\over m}}.
\]
Using the Gauss-Legendre formula  and the
recurrence and reflection formulae for the gamma function
\[
\Gamma(z+1)=z\Gamma(z),\ \ \Gamma(z)\Gamma(1-z)={\pi\over \sin \pi z}
\]
to simplify the coefficients of the series in the last representation
we
obtain the final formula
\begin{equation}
\label{hyper}
\phi({n\over m},\beta;z)  =  \sum\limits_{p=0}^{m-1} {z^p\over
p!\Gamma(\beta+{n\over m}p)}\ _0F_{n+m-1}\left(-;\Delta(n,
{\beta\over n}+{p\over m}),\Delta^*(m,{p+1\over m});{z^m\over
m^mn^n}\right),
\end{equation}
where $ _pF_q((a)_p;(b)_q;z)$ is the generalized
hypergeometric function (\cite[Chapter 4]{Marichev}),
\[
\Delta(k,a) = \{ a,a+{1\over k},\dots,a+{k-1\over k} \},\ \
\Delta^*(k,a) = \Delta(k,a)\setminus \{ 1\}.
\]
We note that the set $\Delta^*(k,a)$ is correctly defined
in our case since the number 1 is an element of the
set $\Delta(m,{p+1\over m}),\ 0\le p \le m-1$.

The same considerations can be applied in the case of negative
rational $\rho$ but under the additional condition that the parameter
$\beta$ is also a rational number. In particular, we obtain the
formulae
\begin{equation}
\label{-1/2.1}
\phi(-{1\over 2},-n;z)={(-1)^{n+1} z\over \pi}\Gamma({3\over 2} +n)\
_1F_1({3\over 2}+n;{3\over 2};-{z^2\over 4}),\ n=0,1,2,\dots,
\end{equation}
\begin{equation}
\label{-1/2.2}
\phi(-{1\over 2},{1\over 2}-n;z)={(-1)^{n} \over \pi}\Gamma({1\over 2} +n)\
_1F_1({1\over 2}+n;{1\over 2};-{z^2\over 4}),\ n=0,1,2,\dots.
\end{equation}
If $n=0$ we get
\begin{equation}
\label{case1}
\phi(-{1\over 2},0;z)=-{z\over 2\sqrt{\pi}}e^{-z^2/4},
\end{equation}
\begin{equation}
\label{case2}
\phi(-{1\over 2},{1\over 2};z)={1\over \sqrt{\pi}}e^{-z^2/4}.
\end{equation}
The formula (\ref{case1}) was given by Stankovi\'c \cite{Sta}. He also gave
the relation ($x>0$)
\[
\phi(-{2\over 3},0;-x^{-{2\over 3}})= -{1\over 2 \sqrt{3\pi}} \exp(-{2\over
27 x^2})W_{-{1\over 2},{1\over 6}}\left(-{4\over 27 x}\right),
\]
where $W_{\mu,\nu}(x)$ is the Whittaker function satisfying the differential
equation
\[
{d^2\over dx^2}W(x) +\left( -{1\over 4} +{\mu\over x} + {\nu^2\over 4 x^2}
\right)W(x) = 0.
\]
The  formula (\ref{case2}) as well as some other particular cases of the Wright
function with $\rho$ and $\beta$ rational, $-1<\rho<0$, can be found in
Mainardi and Tomirotti \cite{MaiTom}, where a particular case
of the Wright function, namely, the function
\begin{equation}
\label{Mai}
M(z;\beta)=\phi(-\beta,
1-\beta;-z),\ 0<\beta<1
\end{equation}
has been considered in details. For $\beta=1/q,\ q=2,3,\dots
$ the representation
\begin{equation}
\label{Main}
M(z;{1\over m})={1\over \pi} \sum_{n=1}^{m-1} (-1)^{n-1}\Gamma(n/m)\sin(\pi
n/m)F(z;n,m)
\end{equation}
with
\[
F(z;n,m)=\sum_{k=0}^\infty (-1)^{k(m+1)}(n/m)_k {z^{mk+n-1}\over
(mk +n-1)!}
\]
was given. Here $(a)_k,\ k=0,1,2,\dots$, denotes the Pochhammer symbol
\[
(a)_k={\Gamma(a+k)\over \Gamma(a)}= a(a+1)\dots(a+k-1).
\]
In particular, the formula (\ref{Main}) gives us for $m=3$
the representation
\[
\phi(-{1\over 3},{2\over 3};z)=3^{2/3}{\mathrm Ai}(-z/3^{1/3})
\]
with the Airy function ${\mathrm Ai}(z)$ .
Finally, we rewrite the formulae (\ref{-1/2.1}), (\ref{-1/2.2}) by
using the Kummer formula (\cite[Chapter 6]{Marichev})
\[
\ _1F_1(a;c;z)=e^z\ _1F_1(c-a;c;-z)
\]
in the form
\begin{equation}
\label{-1/2.3}
\phi(-{1\over 2},-n;z)=e^{-z^2/4}z P_n(z^2), \ n=0,1,2,\dots,
\end{equation}
\begin{equation}
\label{-1/2.4}
\phi(-{1\over 2},{1\over 2}-n;z)=e^{-z^2/4} Q_n(z^2), \ n=0,1,2,\dots,
\end{equation}
where $P_n(z),\ Q_n(z)$ are polynomials of degree $n$ defined as
\[
P_n(z)= {(-1)^{n+1} \over \pi}\Gamma(3/2
+n)\ _1F_1(-n;3/2;{z\over 4}),
\]
\[
Q_n(z)={(-1)^{n} \over \pi}\Gamma(1/2 +n)\ _1F_1(-n;
1/2;{z\over 4}).
\]

\subsection*{2.3. Laplace transform pairs related to the Wright function}

In the case $\rho >0$ the Wright function
is an entire function
of order less than 1 and consequently its Laplace transform can be obtained
by transforming term-by-term its Taylor expansion (\ref{Wright}) in the
origin. As a result we get ($0\le t<+\infty$, $s\in \CC,\
0<\epsilon<|s|,\ \epsilon$ arbitrarily small)
\begin{eqnarray}
\phi(\rho,\beta;\pm t)& \div & {\cal L}[\phi(\rho,\beta;\pm t);s]=
\int_0^\infty e^{-st}\phi(\rho,\beta;\pm t)\, dt \\
 &= & \int_0^\infty e^{-st}\sum_{k=0}^\infty {(\pm t)^k\over k! \Gamma(\rho
k+\beta)}\, dt
=\sum_{k=0}^\infty {(\pm 1)^k\over k! \Gamma(\rho
k+\beta)}\int_0^\infty e^{-st} t^k \, dt \nonumber \\
 &= & {1\over s}\sum_{k=0}^\infty {(\pm s^{-1})^{k} \over \Gamma(\rho
k+\beta)}\ ={1\over s}E_{\rho,\beta}(\pm s^{-1}), \ \ \rho>0, \ \beta \in \CC,
\nonumber
\end{eqnarray}
where $\div$ denotes the juxtaposition of a function $\varphi(t)$ with its
Laplace transform $\tilde \varphi(s)$, and
\begin{equation}
\label{gml}
E_{\alpha,\beta}(z)=\sum_{k=0}^\infty {z^k\over \Gamma(\alpha k+\beta)},\
\alpha>0,\ \beta\in \CC,
\end{equation}
is the generalized Mittag-Leffler function. In this case the resulting
Laplace transform turns out to be analytic, vanishing at infinity and
exhibiting an essential singularity at $s=0$.

For $-1<\rho<0$ the just applied method  cannot be used since then the
Wright function is an entire function of order greater than one.
The existence
of the Laplace transform of the function $\phi(\rho,\beta;-t),\ t>0,$ follows
in this case from Theorem 2.1.3, which says us that the function
$\phi(\rho,\beta;z)$ is exponentially small
for large $z$ in a sector of the plane containing the negative real
semi-axis. To get the transform in this case we use the idea given in Mainardi
\cite{Ma3}. Recalling the integral representation (\ref{repwr}) we have
($-1<\rho<0$)
\begin{eqnarray}
\label{Laplace}
\phi(\rho,\beta;-t) & \div & \int_0^\infty e^{-st}\phi(\rho,\beta;-t)\,
dt = \int_0^\infty e^{-st}{1\over 2\pi i}\int_{{\mathrm Ha}}
e^{\zeta-t \zeta^{-\rho}} \zeta^{-\beta}\, d\zeta \, dt  \nonumber \\
 & = & {1\over 2\pi i}\int_{{\mathrm Ha}}
e^{\zeta}\zeta^{-\beta}\int_0^\infty e^{-t(s+\zeta^{-\rho})}\, dt \, d\zeta
\\
& =  &  {1\over 2\pi i}\int_{{\mathrm Ha}}
{e^{\zeta}\zeta^{-\beta}\over s + \zeta^{-\rho}}\, d\zeta =E_{-\rho, \beta
-\rho}(-s),\nonumber
\end{eqnarray}
again with the generalized Mittag-Leffler function according to
(\ref{gml}). We use here the integral representation
(see Djrbashian \cite{Djr1}, Gorenflo and Mainardi
\cite{GorMai})
\begin{equation}
\label{intgml}
E_{\alpha,\beta}(z)={1\over 2\pi i}\int_{{\mathrm Ha}}
{e^{\zeta}\zeta^{\alpha-\beta}\over \zeta^{\alpha}-z}\, d\zeta,
\end{equation}
which is obtained by substituting the
Hankel representation (\ref{gamma}) for the reciprocal of the
gamma function into the series representation (\ref{gml}).

The relation
(\ref{Laplace}) was given in Djrbashian and Bagian \cite{DjrBag} (see
also Djrbashian \cite{Djr2}) in the case $\beta \ge 0$  as a representation
of the generalized Mittag-Leffler function in the whole complex plane as a
Laplace integral of an entire function and without identifying this
function as the known Wright function. They also gave
(in slightly different notations) the more general representation
\[
E_{\alpha_2,\beta_2}(z)=\int_0^\infty E_{\alpha_1,\beta_1}
(z t^{\alpha_1})t^{\beta_1-1}\phi(-\alpha_2/\alpha_1,\beta_2-\beta_1
{\alpha_2\over \alpha_1};-t)\, dt,
\]
\[
0<\alpha_2<\alpha_1,\ \beta_1,\beta_2>0.
\]
An important particular case of the Laplace transform pair (\ref{Laplace})
is given by
\begin{equation}
\label{LapMai}
M(t;\beta) \div E_\beta(-s),\ 0<\beta<1,
\end{equation}
where $M(t;\beta)$ is the Mainardi function given by (\ref{Mai}) and
\vskip -10pt
\begin{equation}
\label{ml}
E_\alpha(z)=E_{\alpha,1}(z) = \sum_{k=0}^\infty {z^k \over \Gamma(\alpha k
+1)},\ \alpha>0
\end{equation}
is the (standard) Mittag-Leffler function. The formula (\ref{LapMai}) contains,
in particular, the well-known Laplace transform pair
\vskip -8pt
\[
M(t;1/2)={1\over \sqrt{\pi}} \exp(-t^2/4) \div E_{1/2}(-s)= \exp(s^2)\,
\mbox{erfc}\,(s),\ s\in \CC.
\]
Using the relation
\vskip -15pt
\[
\int_0^\infty t^n f(t)\, dt = \lim_{s\to 0} (-1)^n {d^n\over ds^n} {\cal L}
[f(t);s],
\]
the Laplace transform pair (\ref{Laplace}) and the  series representation
of the generalized Mittag-Leffler function (\ref{gml}) we can compute all
the moments of the Wright function $\phi(\rho,\beta;-t),\ -1<\rho<0$ in
$\RR^+$:
\[
\int_0^\infty t^n \phi(\rho,\beta;-t)\, dt ={n!\over \Gamma(-\rho n +
\beta -\rho)},\ n\in \NN_0=\{ 0,1,2,\dots\}.
\]
For the Mainardi function $M(t;\beta),\ 0<\beta<1$ we obtain from this
formula the normalization property in $\RR^+$ ($n=0$)
\vskip -6pt
\[
\int_0^\infty M(t;\beta)\, dt =1
\]
and the moments in the form
\vskip -8pt
\[
\int_0^\infty t^n M(t;\beta)\, dt ={n!\over \Gamma(\beta n+1)},\ n\in \NN.
\]
Now we introduce the function (Mainardi \cite{Ma3})
\vskip -8pt
\begin{equation}
\label{F}
F(z;\beta)=\phi(-\beta,0;-z),\ 0<\beta<1,
\end{equation}
which is connected with the function $M(z;\beta)$ by the relation
\vskip -5pt
\begin{equation}
\label{FM}
F(z;\beta)=\beta z M(z;\beta).
\end{equation}
For this function we can prove the relation
\vskip -5pt
\begin{equation}
\label{FLap}
{1\over t}F(\lambda t^{-\beta};\beta)=
{\beta\lambda \over t^{\beta+1}} M(\lambda t^{-\beta};\beta)
\div \exp(-\lambda s^\beta),\ 0<\beta<1,\ \lambda>0.
\end{equation}
Indeed, following Mainardi \cite{Ma3} and using the integral representation
(\ref{repwr})
we get
\begin{eqnarray}
\label{LaplF}
{\cal L}^{-1}[\exp(-\lambda s^\beta);t] & = & {1\over 2\pi i} \int_{{\mathrm
Ha}}e^{st-\lambda s^{\beta}}ds ={1\over 2\pi i\,t} \int_{{\mathrm Ha}}
e^{\zeta-\lambda t^{-\beta}\zeta^{\beta}}d\zeta  \\
 & = &
{1\over t}F(\lambda t^{-\beta};\beta)=
{\beta\lambda \over t^{\beta+1}} M(\lambda t^{-\beta};\beta). \nonumber
\end{eqnarray}
The Laplace transform pair (\ref{FLap}) was formerly given by Pollard
\cite{Pol} and by Mikusi\'nski \cite{Mik}.

By applying the formula for differentiation of the image  of the Laplace
transform to (\ref{FLap}) we get the Laplace transform pair useful for our
further discussions:
\vskip -6pt
\begin{equation}
\label{MLap}
{1 \over t^{\beta}} M(\lambda t^{-\beta};\beta)
\div s^{\beta -1}\exp(-\lambda s^\beta),\ 0<\beta<1,\ \lambda>0.
\end{equation}

In the general case, using the same method as in (\ref{LaplF}), we get
(see Stankovi\'c \cite{Sta}) the Laplace transform pair
\vskip -6pt
\[
t^{\beta -1}\phi(\rho,\beta;-\lambda t^\rho) \div s^{-\beta} \exp(-\lambda
s^{-\rho}),\ -1<\rho<0,\ \lambda>0.
\]
Stankovi\'c \cite{Sta} also gave some other Laplace transform pairs related
to the Wright function including
\vskip -10pt
\[
t^{{\beta \over 2}-1}\phi(\rho,\beta;-t^{{\rho\over 2}})\div
{\sqrt{\pi}\over 2^\beta}s^{-{\beta\over 2}}\phi({\rho \over 2}, {\beta+1\over 2};
-2^{-\rho}s^{-{\rho\over 2}}),\ \ -1<\rho<0,
\]
\vskip -6pt
\[
t^{-\beta}\exp(-t^{-\rho}\cos(\rho \pi))\sin(\beta \pi - t^{-\rho}\sin(\rho
\pi)) \div \pi s^{\beta -1}\phi (\rho,\beta;-s^\rho),
-1<\rho<0,\ \beta<1.
\]

\subsection*{2.4. The Wright function as an entire function\\
of completely regular growth}

The fact that the function (\ref{Wright}) is an entire function
for all values of the parameters $\rho > -1$ and $\beta \in \CC$
was already known  to Wright
(Wright \cite{Wr1}, \cite{Wright}). In the paper
Djrbashian and Bagian \cite{DjrBag} (see also Djrbashian \cite{Djr2})
the order and type of this function
as well as an estimate of its indicator function were given for the
case $-1<\rho <0$. Wright \cite{Wright}
also remarked that the zeros of the function (\ref{Wright}) lie near
the positive  real semi-axis if $-1/3\le \rho<0$ and near the
two lines $\arg z = \pm{1\over 2} \pi (3\rho +1)$ if $-1 <\rho < -1/3$.
In this paper we continue the investigations of the Wright
function from the viewpoint of the theory of entire functions.
We give exact formulae for the order, the type and the indicator
function of the entire function
$\phi(\rho,\beta;z)$ for $\rho>-1,\ \beta\in \CC$. On the basis of
these results the problem of distribution of zeros of the Wright
function is considered. In all cases this
function is shown to be a function of completely regular growth.

The order and the type of the Wright function are obtained in a
straightforward way by using the standard formulae for the order
$p$ and the type $\sigma$ of an entire function $f(z)$ defined
by the power series $f(z)=\sum_{k=0}^\infty c_n z^n$
\[
p=\limsup_{n\to \infty} {n \log n\over \log(1/|c_n|)},\
(\sigma e p)^{1/p} = \limsup_{n\to \infty}n^{1/p}\sqrt[n]{c_n}
\]
and the Stirling asymptotic  formula
\[
\Gamma(z)=\sqrt{2\pi} z^{z-1/2} e^{-z}[1+O(1/z)],\
|\arg z|\le \pi-\epsilon,\ \epsilon >0,\
|z|\to \infty.
\]
We thus obtain the following result.

\theorem{2.4.1}
The Wright function $\phi(\rho,\beta;z),\ \rho>-1,\
\beta\in \CC\ (\beta\not =-n,\ n=0,1,\dots$ if $\rho =0)$ is
an entire function of finite order  $p$ and the type $\sigma$
given by
\vskip -8pt
\begin{equation}
\label{order}
p={1\over 1+\rho},\ \sigma=(1+\rho)|\rho|^{-{\rho\over 1+\rho}}.
\end{equation}

\vskip 6pt

\remark{2.4.1}
In the case $\rho =0$ the Wright function is reduced to the exponential
function with the constant factor $1/\Gamma(\beta)$,
which turns out to vanish identically for $\beta =-n,\
n=0,1,\dots$. For all other values of the parameter $\beta$ and $\rho =0$
the formulae {\rm (\ref{order})} (with $\sigma =\lim\limits_{\rho \to 0}
(1+\rho)|\rho|^{-{\rho\over 1+\rho}} = 1$) are still valid.

\bigskip

The basic characteristic of the growth of an entire function $f(z)$ of
finite order $p$ in different directions is its indicator function
$h(\theta),\ |\theta|\le \pi$ defined by the equation
\vskip -8pt
\begin{equation}
\label{ind}
h(\theta)=\limsup_{r\to +\infty}{\log |f(re^{
i\theta})|\over r^p}.
\end{equation}
To find the indicator function $h_{\rho}(\theta)$ of the entire
function $\phi(\rho,\beta;z)$ of finite order $p$ given by (\ref{order})
its asymptotics given in Section 2.1 are used. By direct evaluations
we arrive at the following theorem.

\theorem{2.4.2}
Let $\rho>-1,\ \beta\in \CC\ (\beta\not
=-n,\ n=0,1,\dots$ if $\rho =0)$. Then the indicator function
$h_{\rho}(\theta)$ of
the Wright function $\phi(\rho,\beta;z)$ is given by the formulae
\begin{equation}
\label{ind1}
h_{\rho}(\theta) = \sigma \cos p\theta,\ |\theta|\le \pi
\end{equation}
in the case $\rho \ge 0$,
\begin{equation}
\label{ind2}
h_{\rho}(\theta) = \cases{
-\sigma \cos p( \pi+\theta), & $-\pi\le \theta \le 0$, \cr
-\sigma \cos p(\theta-\pi),& $0\le
\theta\le \pi $\cr}
\end{equation}
in the cases (a) $ -1/3\le \rho<0$, (b) $\rho =-1/2,\ \beta=-n,\
n=0,1,\dots$ and (c) $\rho=-1/2,\ \beta=1/2-n,\ n=0,1,\dots$,
\begin{equation}
\label{ind3}
h_{\rho}(\theta) = \cases{-\sigma \cos p( \pi+\theta), &
$-\pi\le \theta \le {3\over 2}{\pi\over p} -\pi$, \cr
0,& $|\theta| \le \pi- {3\over 2}{\pi\over p}$, \cr
-\sigma \cos
p(\theta-\pi),& $ \pi- {3\over 2}{\pi\over p}\le \theta\le \pi$ \cr}
\end{equation}
in the case $ -1< \rho<-1/3$ $(\beta\not = -n,\ n=0,1,\dots$ and
$\beta\not = 1/2-n,\ n=0,1,\dots$ if $\rho=-1/2)$,
where $p$ and  $\sigma$ are the order and type of the Wright
function, respectively, given by {\rm (\ref{order})}.

\remark{2.4.2}
It can be seen from the formulae {\rm (\ref{ind1}), (\ref{ind2})} that
the indicator function $h_{\rho}(\theta)$ of the Wright function
$\phi(\rho,\beta;z)$ is reduced to the function $\cos \theta$ -- the
indicator function of the exponential function $e^z$ -- if $\rho\to
0$. This property is not valid for another generalization of the
exponential function -- the Mittag-Leffler function {\rm (\ref{ml})}.
Even though
\[
E_{1}(z)=e^z,
\]
the indicator function of the Mittag-Leffler function given for
$0<\alpha<2,\ \alpha\not =1$ by {\rm (\cite[Chapter 2.7]{Evg})}
\[
h(\theta)=\cases{ \cos\theta/\alpha,& $|\theta|\le{\pi \alpha\over 2}$,\cr
                  0,& ${\pi \alpha\over 2}\le |\theta|\le \pi$ \cr}
\]
does not coincide with the indicator function of $e^z$ if $\alpha\to 1$.

\bigskip

We consider now the problem of distribution of zeros of the Wright function
in the case $\rho >-1,\ \beta\in \RR$.
To get the asymptotics of zeros of the Wright function
we use its asymptotic expansions (\ref{a0_1}), (\ref{a0_2}), (\ref{a1_1}),
(\ref{a1_2}), (\ref{a1_3}), (\ref{a2_1})
and the method applied by M.M. Djrbashian in \cite[Chapter
1.2]{Djr1} to solve the problem of distribution of zeros of the
ge\-ne\-ra\-lized Mittag-Leffler function $E_{\rho,\mu}(z)$. This method
consists in finding the asymptotics of zeros of the main terms of the
asymptotic expansions, applying the
Rouch\'e theorem to show that the function under consideration and the main terms of its
asymptotic expansions have the same number of zeros inside of specially
chosen contours and after that in estimation of the diameter of the
domains bounded by the contours. The proofs of the results given below are
straightforward but have many technical details and  are omitted in
this paper.
It turns out, that
in dependence of the value of the parameter $\rho>-1$ and the real
parameter $\beta$, there are five different situations:

1) for $\rho >0$
all zeros with large enough absolute values are simple and are lying on the
negative real semi-axis;

2) in the case $\rho=0$ the Wright function becomes
the exponential function with a constant factor (equal
to zero if $\beta=-n,\ n=0,1,\dots$) and it has no zeros;

3) for $-1/3\le
\rho<0$
 all zeros with large enough absolute values are simple and are lying on the
positive real semi-axis;

4) in the cases $\rho=-1/2,\ \beta=-n,\
n=0,1,\dots$ and
$\rho=-1/2,\ \beta=1/2-n,\ n=0,1,\dots$ the Wright function has exactly
$2n+1$ and $2n$ zeros, respectively;

5) for $-1<\rho<-1/3$ (excluding the
case 4)) all  zeros with large enough absolute values are simple
and are lying in the neighbourhoods of the rays $\arg z=\pm
{1\over 2}\pi (-1-3\rho)$.

\bigskip

We now give the precise results.

\theorem{2.4.3}
Let $\{\gamma_k\}_1^\infty$ be the sequence of zeros of the function
$\phi(\rho,\beta;z)$, $\rho\ge -1/3,\ \rho\not =0,\
\beta \in \RR $, where $|\gamma_k|\le |\gamma_{k+1}|$
and each zero is counted according to its multiplicity.  Then:

{\em {\bf A}. In the case $\rho >0$ all zeros with large
enough $k$ are simple and are lying on the negative real semi-axis. The
asymptotic formula
\begin{equation}
\label{zero1}
\gamma_k=-\left( {\pi k +\pi( p \beta -{p-1\over 2})\over \sigma \sin \pi
p}\right)^{{1\over p}}\left\{ 1 +O(k^{-2})\right\},\ k\to +\infty
\end{equation}
is true. Here and in the next formulae $p$ and
$\sigma$ are the order and type of the Wright function given by
{\rm (\ref{order})}, respectively.

{\bf B}. In the case $-1/3\le \rho <0$ all zeros with large
enough $k$ are simple, lying on the positive real semi-axis and the
asymptotic formula
\begin{equation}
\label{zero2}
\gamma_k=\left({ \pi k +\pi( p \beta -{p-1\over 2})\over -\sigma \sin \pi
p}\right)^{{1\over p}}\left\{ 1 +O(k^{-2})\right\},\ k\to +\infty
\end{equation}
is true.}

\remark{2.4.3}
Combining the representation {\rm (\ref{Bess})} with the asymptotic formula
{\rm (\ref{zero1})} we get the known formula (see, for example {\rm
\cite[p.506]{Wat}}) for asymptotic expansion of the large zeros $r_k$ of
the Bessel function $J_\nu (z)$:
\[
r_k = \pi(k+{1\over 2}\nu -{1\over 4}) +O(k^{-1}),\ k\to \infty.
\]

\remark{2.4.4}
In the cases $\rho=-1/2,\ \beta=-n,\
n=0,1,\dots$ and
$\rho=-1/2,\ \beta=1/2-n,\ n=0,1,\dots$ the Wright function  can be
represented by the formulae {\rm (\ref{-1/2.3})},  {\rm (\ref{-1/2.4})} and,
consequently, has exactly $2n+1$ and $2n$ zeros in the complex plane,
respectively.

\bigskip

It follows from the asymptotic formulae (\ref{a1}), (\ref{a2})
and (\ref{a2_1}) that all zeros of the function
$\phi(\rho,\beta;z)$ in the case $-1<\rho <-1/3$
with large enough absolute values are lying inside of the angular
domains
\[
\Omega_\epsilon^{(\pm)}=\left\{ z:\ \left| \arg z \mp \left( \pi -{3\pi \over 2 p}\right)
\right| <\epsilon\right\},
\]
where $\epsilon$ is any number of the interval $(0,\min\{
\pi -{3\pi \over 2 p},{3\pi \over 2 p}\})$. Consequently, the function
$\phi(\rho,\beta;z)$ has on the real axis only finitely many zeros.
Let
\[
\{ \gamma_k^{(+)}\}_1^\infty\in G^{(+)} =\{z:\ \Im >0 \},\
\{ \gamma_k^{(-)}\}_1^\infty\in G^{(-)} =\{z:\ \Im <0 \}
\]
be sequences of zeros of the function $\phi(\rho,\beta;z)$ in the
upper and lower half-plane, respectively, such that
$|\gamma^{(+)}_k|\le |\gamma^{(+)}_{k+1}|$,
$|\gamma^{(-)}_k|\le |\gamma^{(-)}_{k+1}|$,
and each zero is counted according to its multiplicity.

\theorem{2.4.4}
In the case $-1<\rho <-1/3$ $(\beta\not = -n,\ n=0,1,\dots$ and
$\beta\not = 1/2-n,\ n=0,1,\dots$ if $\rho=-1/2)$ all zeros
of the function
$\phi(\rho,\beta;z),\ \beta\in \RR$
with large
enough $k$ are simple and the asymptotic formula
\begin{equation}
\label{zero3}
\gamma^{(\pm)}_k = e^{\pm i (\pi -{3\pi \over 2 p})}
\left( {2\pi k \over \sigma}\right)^{{1\over p}}
\left\{ 1 +O\left( {\log k\over k}\right)\right\},\ k\to +\infty
\end{equation}
is true.

\bigskip

Summarizing all results concerning the asymptotic behaviour
of the Wright function, its indicator function and the distribution
of its zeros, we get the theorem.

\theorem{2.4.5}
The Wright function $\phi(\rho,\beta;z),\ \rho >-1$ is an entire
function of completely regular growth.

\bigskip

We recall (\cite[Chapter 3]{Lev}) that an entire  function $f(z)$ of
finite order $p$ is called a function of completely regular growth
(CRG-function) if for all $\theta,\ |\theta|\le \pi$, there exist
a set $E_\theta \subset \RR_+$ and  the limit
\vskip -10pt
\begin{equation}
\lim_{{r\to +\infty\atop r\in E_\theta^* }} {\log
\mid f(re^{i\theta}
)\mid\over r^p},
\label{44}
\end{equation}
\vskip -3pt \noindent
where
\vskip -8pt
\[
E_\theta^* = \RR_+ \setminus E_\theta,\ \ \lim_{r\to +\infty} {{\rm
mes}E_\theta\bigcap(0,r)\over r} = 0.
\]
It is known (\cite[Chapter 2.6]{Evg})
that  zeros of a CRG-function $f(z)$
are regularly distributed, namely,
they possess the finite  angular
density
\begin{equation}
\lim_{r\to +\infty} {n(r,\theta)\over r^p} = \nu(\theta),
\label{45}
\end{equation}
where $n(r,\theta)$ is the number of zeros of $f(z)$ in the sector $0<\arg
z < \theta,\ |z|<r$ and $p$ is the order of $f(z)$.
>From the other side, the angular density $\nu(\theta)$ is connected with
the indicator function $h(\theta)$ of a CRG-function. In particular (see
\cite[Chapter 2.6]{Evg}),
the jump of $h'(\theta)$ at $\theta=\theta_0$ is equal to $2\pi p \Delta$,
where $\Delta$ is the density of zeros of $f(z)$ in an arbitrarily small
angle containing the ray $\arg z =\theta_0$.

In our case
we get from Theorem 2.4.2, that the derivative of the indicator
function of the Wright function has the jump $2\sigma p \sin \pi p$ at
$\theta=\pi$ for $\rho >0$, the same jump at $\theta=0$ for $-1/3<\rho<0$,
and the jump $\sigma p$ at $\theta=\pm (\pi -{3\pi \over 2 p})$ for
$ -1< \rho<-1/3$ ($\beta\not = -n,\ n=0,1,\dots$ and
$\beta\not = 1/2-n,\ n=0,1,\dots$ if $\rho=-1/2$),
where again $p$ and  $\sigma$ are the order and type of the Wright
function, respectively, given by (\ref{order}); if $\rho =0$ or $\rho
=-1/2$  and either $\beta= -n,\ n=0,1,\dots$, or
$\beta = 1/2-n,\ n=0,1,\dots$, the derivative of the indicator function has
no jumps. As we see, the behaviour of the derivative of the indicator
function of the Wright function is in accordance with  the distribution of
its zeros given by Theorems 2.4.3, 2.4.4 and Remark 2.4.4 as predicted by
the general theory of the CRG-functions.

\section*{3. Some applications of the Wright function}

\subsection*{3.1. Asymptotic theory of partitions}

Historically the first application of the Wright function was
connected with the asymptotic theory of partitions. Extending the results
of Hardy and Ramanujan about asymptotic expansion of the function
$p(n),\ n\in \NN$, the number of partitions of $n$, Wright \cite{Wri1}
considered the more general problem, namely, to find an asymptotic expansion
for the function $p_k(n),\ n\in \NN$, the number of partitions of $n$ into
perfect $k$-th powers. Following Hardy and Ramanujan, Wright considered
the generating function for the sequence $\{p_k(1),\ p_k(2),\dots\}$
which is given by
\vskip -10pt
\[
f_k(z) = \prod_{l=1}^\infty (1-z^{l^k})^{-1} = 1+\sum_{n=1}^\infty p_k(n)z^n,
\ |z|<1.
\]
\vskip -3pt \noindent
Then
\vskip -15pt
\[
p_k(n) ={1\over 2\pi i} \int_{{\rm C}} {f_k(z)\, dz\over z^{n+1}},
\]
the contour ${\rm C}$ being the periphery  of the circle with
center in the point $z=0$ and  radius $r=1-{1\over n}$. Let the contour
be divided into a large number of small arcs, each associated with a point
\vskip -14pt
\[
\alpha_{p,q} =\exp (2p\pi i/q),\ p,q\in \NN.
\]
Taking the arc associated with $\alpha_{0,1}=1$ as typical, it can be shown
that on this arc the generating function $f_k(z)$ has the representation
\vskip -6pt
\begin{equation}
\label{fk}
f_k(z) \sim {z^j \over (2\pi)^{{1\over 2}(k+1)}}\left( \log {1\over z}
\right)^{{1\over 2}} \exp \left( {\Gamma(1+(1/k))\zeta(1+(1/k))
\over (\log(1/z))^{1/k}}\right),\ z\to 1,
\end{equation}
\vskip -2pt \noindent
where $j$ is a real number depending on $k$ and $\zeta(z)$ is the Riemann
zeta-function. Then, on this arc, $f_k(z)$ is approximated to by an
auxiliary function $F_k(z)$, which has a singularity at $z=1$ of the type
of the right-hand side of (\ref{fk}). If the $z$-plane is cut along the
interval $(1,\infty)$ of the real axis, $F_k(z)$ is regular and one-valued
for all values of $z$ except those on the cut. The power series for $F_k(z)$
has coefficients given in terms of the entire function $\phi(\rho,\beta;z)$
and, by using this power series, an asymptotic expansion can be found for
$p_k(n)$.

In the paper \cite{Wri1} Wright gave some properties of the function
$\phi(\rho,\beta;z)$ in the case $\rho >0$, including its asymptotics and
integral representation (\ref{repwr}). He proved on this base the following
two theorems.

\theorem{3.1.1}
Let $\alpha,\ \beta,\ \gamma \in \CC,\ \alpha\not =0,\ \rho >0,\ m\in \NN,
\ m>\Re(\gamma)$,
\begin{equation}
\label{F(x)}
F(z) = F(\rho,\alpha,\beta,\gamma;z):= \sum_{n=m}^\infty (n-\gamma)^{\beta-1}
\phi(\rho,\beta;\alpha(n-\gamma)^\rho)z^n.
\end{equation}
If a cut is made in the $z$-plane along the segment $(1,\infty)$ of the
real axis, then $F(z)$ is regular and one-valued in the interior of the
region thus defined.

\theorem{3.1.2}
Let
\[
G(z)=F(z)-\chi(z),
\]
where $F(z)$ is defined by {\rm (\ref{F(x)})} and
\[
\chi(z)={z^\gamma \over (\log(1/z))^\beta} \exp \left({\alpha\over
(\log(1/z))^\rho}\right).
\]
If a cut is made in the $z$-plane along the segment $(-\infty,0)$ of
the real axis, then $G(z)$ is regular and one-valued in the interiour
of the region thus defined.

\bigskip

We see that the function $F(x)$ has a singularity of the type of
$\chi(z)$ at $z=1$.  In the case of the function $F_k(z)$ used to get
an asymptotic expansion for the function $p_k(n)$ the values
\vskip -6pt
\[
\rho={1\over k},\ \alpha=\Gamma(1+{1\over k})\zeta(1+{1\over k}),\
\beta =-{1\over 2},\ \gamma ={1\over 24}
\]
should be taken in the previous two theorems.

\subsection*{3.2. Fractional diffusion-wave equation}

Another field in which the Wright function plays a very important role
is that of partial differential equations of fractional order. Following
Gorenflo, Mainardi and Srivastava \cite{GorMaiSri} and Mainardi
\cite{Ma1}--\cite{Ma3}
we consider  the fractional diffusion-wave equation which is
obtained from the classical diffusion or wave equation by replacing the
first- or second-order time derivative by a fractional derivative of order
$\alpha$ with $0<\alpha\le 2$:
\begin{equation}
\label{FPDE}
\FTS{\partial^{\alpha}u(x,t)}{ \partial t^{\alpha}}={\cal
D} \FTS{\partial^{2}u(x,t)}{ \partial x^{2}}, \ \  {\cal D}>0,\ \
0<\alpha\le 2.
\end{equation}
Here the field variable $u=u(x,t)$ is assumed to be a causal function of
time, i.e. vanishing for $t<0$, and the fractional derivative is taken in
the Caputo sense:
\begin{equation}
\label{CFD}
\FTS{\partial^{\alpha}u(x,t)}{\partial
t^{\alpha}}=\cases{
\FTS{\partial^{n}u(x,t)}{\partial t^{n}},&$\alpha=n \in\NN$, \cr
\FTS{1}{\Gamma(n-\alpha)}\int\limits^{t}_{0}(t-\tau)^{n-\alpha-1}
\FTS{\partial^n
u(x,\tau)}{\partial \tau^n}\, d\tau, &$n-1<\alpha<n$. \cr}
\end{equation}
We refer to the equation
(\ref{FPDE})  as to the {\em fractional diffusion\/}
and to the {\em fractional wave\/} equation in the cases
$0<\alpha\le 1$ and $1<\alpha\le 2$, respectively. The
difference between these two cases can be seen in the
formula for the Laplace transform of the Caputo fractional
derivative of order $\alpha$  (see Mainardi \cite{Ma3}):
\vskip -6pt
\begin{equation}
\label{CFDLap}
\FTS{\partial^{\alpha}u(x,t)}{\partial t^{\alpha}}\div s^\alpha \tilde u
(x,s) -\sum_{k=0}^{n-1} s^{\alpha-1-k}
\FTS{\partial^{k}u(x,t)}{\partial t^{k}}\vert_{t=0+},\
n-1<\alpha\le n,\ n\in \NN.
\end{equation}

Extending the conventional analysis to the equation (\ref{FPDE}), and
denoting by $g(x)$ and $h(x)$ two given, sufficiently well-behaved
functions, the basic boundary-value problems can be formulated
as follows ($0<\alpha\le 1$):

\bigskip

a) Cauchy problem
\begin{equation}
\label{Cauchy}
u(x,0+)=g(x),\ -\infty<x<+\infty;\ \ u(\mp \infty,t)=0,\ t>0;
\end{equation}

b) Signalling problem
\begin{equation}
\label{Sign}
u(x,0+)=0,\ x>0;\ \ u(0+,t)=h(t),\ u(+\infty,t)=0,\ t>0.
\end{equation}
If $1<\alpha\le 2$ the initial values of the first time-derivative
of the field variable, $\dot u(x,0+)$, should be added to to
the conditions (\ref{Cauchy}) and (\ref{Sign}). To ensure the continuous
dependence of the solutions on the parameter $\alpha$ in the transition
from $\alpha=1-$ to $\alpha=1+$, we agree to assume $\dot u(x,0+)=0$.

Since these problems are well studied in the cases $\alpha=1$ and
$\alpha=2$ we restrict ourselves in the further considerations to
the case $0<\alpha<2,\ \alpha\not = 1$. For the sake of convenience we
use the abbreviation
\vskip -6pt
\begin{equation}
\label{beta}
\beta = {\alpha\over 2},
\end{equation}
\vskip -2pt \noindent
which implies $0<\beta<1$.

Let us introduce the Green functions ${\cal G}_c(x,t;\beta)$ and
${\cal G}_s(x,t;\beta)$ for the Cauchy and signalling problems for
the equation (\ref{FPDE}),
respectively, which represent the fundamental solutions of these
problems (with $g(x)=\delta(x)$ in (\ref{Cauchy}) and $h(t)=\delta(t)$
in (\ref{Sign})). Using the Green functions, the solutions of the
two basic problems can be given, respectively, by
\begin{equation}
\label{Csol}
u(x,t;\beta)=\int_{-\infty}^{+\infty} {\cal G}_c(x-\xi,t;\beta)g(\xi)
\, d\xi,
\end{equation}
\begin{equation}
\label{Ssol}
u(x,t;\beta)=\int_{0}^{t} {\cal G}_s(x,t-\tau;\beta)h(\tau)
\, d\tau.
\end{equation}
To get the Green functions ${\cal G}_c(x,t;\beta)$ and
${\cal G}_s(x,t;\beta)$ the technique of the Laplace transform is used.
We consider at first the Cauchy problem (\ref{Cauchy}) for the equation
(\ref{FPDE}) with $g(x)={\cal G}_c(x,0+;\beta)=\delta(x)$ (and $\dot {\cal
G}_c(x,0+;\beta)=0$ if $1/2<\beta<1$). Denoting the Laplace transform
of the Green function by $\tilde {\cal G}_c(x,s;\beta)$ and using
the formula (\ref{CFDLap}) we arrive
after application of the Laplace transform to the Cauchy problem
\{(\ref{FPDE}), (\ref{Cauchy})\} to the non-homogeneous
differential equation
\vskip -10pt
\begin{equation}
\label{dif1}
{\cal D}{d^2 \tilde {\cal G}_c \over dx^2} -s^{2\beta} \tilde {\cal G}_c
=-\delta(x) s^{2\beta -1},\ -\infty <x<+\infty
\end{equation}
with the boundary conditions
\vskip -10pt
\begin{equation}
\label{c1}
\tilde {\cal G}_c(\mp \infty,s;\beta) =0.
\end{equation}
The problem \{(\ref{dif1}), (\ref{c1})\} has a solution (see, for
example, Mainardi \cite{Ma3})
\begin{equation}
\label{LapGC}
\tilde {\cal G}_c(x,s;\beta) ={1\over 2\sqrt{{\cal D}}\, s^{1-\beta}}
e^{-(|x|/\sqrt{D})s^\beta},\ -\infty <x<+\infty.
\end{equation}
Comparing this relation with the Laplace transform pair (\ref{MLap})
we represent the Green function for the Cauchy problem
\{(\ref{FPDE}), (\ref{Cauchy})\} in the form
\begin{equation}
\label{GrC}
{\cal G}_c(x,t;\beta)={r\over 2\sqrt{{\cal D}}|x|}M(r/\sqrt{{\cal
D}};\beta),\ t>0,
\end{equation}
where
\vskip -14pt
\[
r=|x|\, t^{-\beta}
\]
is the {\em similarity variable\/} and $M(z;\beta)$ is the Mainardi
function (\ref{Mai}) given in terms of the Wright function.

For the signalling problem \{(\ref{FPDE}), (\ref{Sign})\}
(with $h(t)=\delta(t)$) the
application of the Laplace transform leads to the homogeneous
differential equation
\vskip -8pt
\begin{equation}
\label{dif2}
{\cal D}{d^2 \tilde {\cal G}_s \over dx^2} -s^{2\beta} \tilde {\cal G}_s
=0,\ x\ge 0
\end{equation}
with the boundary conditions
\vskip -14pt
\begin{equation}
\label{c2}
\tilde {\cal G}_s(0+,s;\beta) =1,\ \tilde {\cal G}_s(+\infty,s;\beta)=0.
\end{equation}
Solving this equation, we obtain
\vskip -14pt
\begin{equation}
\label{LapGS}
\tilde {\cal G}_s(x,s;\beta) =
e^{-(x/\sqrt{D})s^\beta},\ x\ge 0.
\end{equation}
Using the Laplace transform pair (\ref{FLap}) we get the Green function
${\cal G}_s(x,t;\beta)$ for the signalling problem
\{(\ref{FPDE}), (\ref{Sign})\} in the form
\vskip -6pt
\begin{equation}
\label{GrS}
{\cal G}_s(x,t;\beta)={\beta r\over \sqrt{{\cal D}}
t}M(r/\sqrt{{\cal D}};\beta),\ t>0,\ x\ge 0,
\end{equation}
where
\vskip -14pt
\begin{equation}
\label{simvar}
r=x\, t^{-\beta}
\end{equation}
is the {\em similarity variable} and $M(z;\beta)$ is the Mainardi
function (\ref{Mai}).

\bigskip

For more results in FPDE we refer, for example,  to
Engler \cite{Eng1}, Fujita \cite{Fu1}, Gorenflo and Mainardi
\cite{Go1}, \cite{Go2}, Mainardi \cite{Ma1}-\cite{Ma3}, Podlubny
\cite{Pod}, Pr\"uss \cite{Pr1}, Saichev and Zaslavsky \cite{Sai1},
Samko et. al. \cite{Samko}, Schneider and
Wyss \cite{Sc1} and by Wyss \cite{Wy1}.

Some applications of FPDE have been considered in papers by
several authors including Giona and Roman \cite{GionaRoman92},
Hilfer \cite{Hilfer95}, Mainardi \cite{Mai},
Metzler et al. \cite{Metzler94}, Nigmatullin \cite{Nig}, Pipkin \cite{Pip},
Podlubny \cite{Pod}.

\subsection*{3.3. Scale invariant solutions of FPDE}

Let us consider the abstract equation
\vskip -6pt
\begin{equation}
\label{abstreq}
F(u)\ =\ 0,\ \ u=u(x,t).
\end{equation}

First we  give some definitions concerning the
similarity method.

\definition{3.3.1}
A one-parameter family of scaling transformations,
denoted by $T_{\lambda}$, is a transformation of $(x,t,u)$-space
of the form
\vskip -5pt
\begin{equation}
\label{trans1}
\bar{x} \ = \ \lambda^{a} x, \quad
\bar{t} \ = \ \lambda^{b} t, \quad
\bar{u} \ = \  \lambda^{c} u,
\end{equation}
\vskip -2pt \noindent
where $a,$ $b$, and $c$ are constants and $\lambda$ is a real
parameter restricted to an open interval $I$ containing $\lambda = 1$.

\definition{3.3.2}
The equation (\ref{abstreq}) is invariant under
the one-parameter family $T_{\lambda}$ of scaling transformations
(\ref{trans1}) iff $T_{\lambda}$ takes any solution $u$ of (\ref{abstreq})
to a solution $\bar{u}$ of the same equation:
\vskip -14pt
\begin{equation}
\label{transeq}
\bar{u} \ =\ T_{\lambda}u \quad \mbox{ and } \quad F(\bar{u})\ =\ 0.
\end{equation}

\definition{3.3.3}
A real-valued function $\eta(x,t,u)$ is called an
{\it invariant\/} of the one-parameter family $T_{\lambda}$, if it is
unaffected by the transformations, in other words:
\vskip -12pt
\[ \eta(T_{\lambda}(x,t,u))\ =\ \eta(x,t,u) \quad \mbox{for all}
 \quad \lambda \in I.\]

On the half-space $\{(x,t,u):x>0,\ t>0\}$, the invariants of the family
of scaling transformations (\ref{trans1}) are provided by the functions
(see \cite{Olver})
\begin{equation}
\label{eta}
\eta_1(x,t,u)=xt^{-a/b},\ \eta_2(x,t,u)=t^{-c/b}u.
\end{equation}
If the equation (\ref{abstreq}) is a second order partial differential
equation of the form
\begin{equation} \label{pde}
G(x,\ t,\ u,\ u_x,\ u_t,\ u_{xx},\ u_{tt},\ u_{xt}) \ = \ 0,
\end{equation}
and this equation is invariant under $T_{\lambda}$, given by
(\ref{trans1}), then the transformation
\begin{equation}
\label{simtrans}
u(x,t) \ = \  t^{c/b}v(z), \quad z \ = \ xt^{-a/b}
\end{equation}
reduces the equation (\ref{pde}) to a second order ordinary
differential equation of the form
\vskip -6pt
\begin{equation} \label{ode}
g(z,\ v,\ v',\ v'') \ = \ 0.
\end{equation}
For a proof of this fact we refer in the case of general
Lie group methods to \cite{Olver}.  In some
cases it can be easily checked directly.

Recently, the scale-invariant solutions for the equation (\ref{FPDE}) (with
the fractional derivative in the Caputo and Riemann-Liouville sense) and for
the more general time- and space-fractional partial differential equation
(with the Riemann-Liouville space-fractional derivative of  order
$\beta\le 2$ instead of the second order space derivative in the
equation (\ref{FPDE}))  have been obtained by Gorenflo, Luchko and
Mainardi \cite{GorLucMai}, Buckwar and Luchko \cite{BucLuc} and Luchko
and Gorenflo \cite{Luc}, respectively.
In all cases these solutions have been given in terms of the Wright
and the generalized Wright functions. Here we present some results from
these papers.

At first we determine a group of scaling transformations for the
fractional diffusion-wave equation (\ref{FPDE}) on the semi-axis ($x\ge 0$)
with the Caputo fractional derivative given by (\ref{CFD}). We have in this
case the following theorem.

\theorem{3.3.1}
\label{invar}
Let
$T_\lambda$ be  a one parameter group of scaling transformations for
the equation (\ref{FPDE}) of the form $T_\lambda \circ (x,t,u)=(\lambda
x,\ \lambda^b t,\ \lambda^c u)$.
Then,
\begin{equation}
\label{b}
b={2\over \alpha}
\end{equation}
and the invariants  of this
group $T_\lambda$ are given
by the expressions
\begin{equation}
\label{inv}
\eta_1(x,t)=xt^{-1/b}=xt^{-\alpha/2},\
\eta_2(x,t,u)=t^{-c/b}u=t^{-\gamma}u
\end{equation}
with a real parameter $\gamma=c \alpha /2$.

\smallskip

\remark{3.3.1}
We note that the first scale-invariant $\eta_1$ of
{\em (\ref{inv})} coincides with the similarity variable
{\em (\ref{simvar})} which was used to define
the Green function of the signalling boundary-value
problem for the equation {\em (\ref{FPDE})}. It is a consequence
of the fact that the equation {\em (\ref{FPDE})} is invariant under
the corresponding group of scaling transformations.

\bigskip

It follows from the general theory of Lie groups and the previous
theorem that the scale-invariant solutions of the equation (\ref{FPDE})
should have the form
\vskip -5pt
\begin{equation}
\label{v}
u(x,t)=t^\gamma v(y),\ y=x t^{-\alpha/2}.
\end{equation}
Furthermore, the general theory says that the substitution (\ref{v})
reduces the
partial integro-differential equation (\ref{FPDE}) into an ordinary
integro-differential equation with the unknown function $v(y)$.

\theorem{3.3.2}
The reduced equation for the scale-invariant solutions of the equation
{\rm (\ref{FPDE})} of the form {\rm (\ref{v})} is given by
\begin{equation}
\label{red}
( _*P_{2/\alpha}^{\gamma-n+1,\alpha}v)(y)={\cal D}v''(y),\ y>0,
\end{equation}
where the operator in the left-hand side is the Caputo type
modification of the left-hand sided Erd\'elyi-Kober fractional
differential operator defined for
$ 0<\delta,\ n-1<\alpha\le n\in \NN$ by
\vskip -10pt
\begin{equation}
( _*P_\delta^{\tau,\alpha}g)(y):=(K_\delta^{\tau,n-\alpha}
\prod_{j=0}^{n -1} (\tau +j-{1\over
\delta} u{d\over du})g)(y),\ y>0.
\label{C}
\end{equation}
\vskip -4pt \noindent
Here
\vskip -15pt
\begin{equation}
\label{K}
(K_\delta^{\tau,\alpha} g)(y):=
\left\{ \begin{array}{ll} {1\over \Gamma(\alpha)}
\int_1^\infty (u-1)^{\alpha -1} u ^{-(\tau+\alpha)}
g(yu^{1/\delta})\, du, & \alpha>0,\\ [6pt]
g(y),& \alpha=0
\end{array}
\right.
\end{equation}
is the left-hand sided Erd\'elyi-Kober fractional integral operator.

\smallskip

\remark{3.3.2}
As it follows from the definitions of the Caputo type
modification of the Erd\'elyi-Kober fractional differential operator
{\rm (\ref{C})} and the Erd\'elyi-Kober
fractional integral operator {\rm (\ref{K})}  in the case
$\alpha =n \in \NN$, the equation {\rm (\ref{red})}
for the scale-invariant solutions is a linear ordinary differential
equation of order $\max\{n,2\}$. In the case $\alpha=1$ (the
diffusion equation) we have
\vskip -6pt
\[
( _*P_2^{\gamma,1} v)(y) = (\gamma -{1\over 2} y{d\over dy})v(y)
\]
and {\rm (\ref{red})} takes the form
\vskip -12pt
\begin{equation}
{\cal D}v''(z)+{1\over 2}y v'(y)-\gamma v(y)=0.
\end{equation}
In the case $\alpha=2$ (the wave equation) we get
\vskip -5pt
\begin{eqnarray}
( _*P_1^{\gamma-1,2} v)(y) & = & (\gamma -1- y{d\over dy})
(\gamma -y{d\over dy})v(y)
\nonumber \\
& = & y^2v''(y)-2(\gamma-1)yv'(y) +\gamma(\gamma-1)v(y)
\nonumber
\end{eqnarray}
and {\rm (\ref{red})} is reduced to the ordinary differential equation
of the second order:
\vskip -4pt
\begin{equation}
(y^2-{\cal D})v''(y)-2(\gamma-1)yv'(y) +\gamma(\gamma-1)v(y) =0.
\end{equation}
The complete discussion of these cases one can find, for example,
in {\rm \cite{Olver}}. The case
$\alpha =n \in \NN,\ n>2$ was considered in {\rm \cite{BucLuc}}.

\bigskip

Solving the equation (\ref{C}) we get the following theorems.

\theorem{3.3.3}
The scale-invariant solutions of the fractional diffusion
equation {\rm (\ref{FPDE})} $(0<\alpha \le 1)$  have the form
\vskip -7pt
\begin{equation}
\label{solA}
u(x,t)=C_1t^\gamma \phi(-{\alpha\over 2},1+\gamma;
-{y\over \sqrt{{\cal D}}})
\end{equation}
\vskip -3pt \noindent
in the case $-1<\gamma,\ \gamma\not=0$, and
\vskip -14pt
\begin{equation}
\label{solB}
u(x,t) = C_1\phi(-{\alpha\over 2},1;-{y\over \sqrt{{\cal
D}}}) +C_2
\end{equation}
\vskip -3pt \noindent
in the case $\gamma=0$, where $y=xt^{-{\alpha\over 2}}$ is the
first scale invariant {\rm (\ref{inv})} and
$C_1,  C_2$ are arbitrary constants.

\theorem{3.3.4}
The scale-invariant solutions of the fractional wave
equation {\rm (\ref{FPDE})} $(1<\alpha<2)$  have the form
\vskip -4pt
\begin{eqnarray}
\label{solC}
u(x,t) & = & C_1 t^\gamma
\phi(-{\alpha\over 2},1+\gamma;-{y\over \sqrt{{\cal
D}}})
\\
& + & C_2t^\gamma
\biggl( {{\cal D}^{{\gamma-1\over \alpha}}
\over 2}
\phi(-{\alpha\over 2},1+\gamma;{y\over  \sqrt{{\cal
D}}})
\nonumber \\
& - & {y^{2+2{\gamma -1\over \alpha}} \over {\cal D}}
\phi((-\alpha,2-\alpha),(2,3+2{\gamma-1\over \alpha});
{y^2\over {\cal
D}})\biggr),
\nonumber
\end{eqnarray}
in the case $1-\alpha<\gamma <1,\ \gamma\not =1-{\alpha\over 2},\
\gamma\not=0$, and
\begin{equation}
\label{solD}
u(x,t) = C_1\phi(-{\alpha\over 2},1;-{y\over \sqrt{{\cal D}}})
\end{equation}
\[
+C_2\left( {{\cal D}^{-{1\over \alpha}}\over 2}
\phi(-{\alpha\over 2},1;{y\over \sqrt{{\cal
D}}}) - {y^{2-{2\over \alpha}} \over {\cal D}}
\phi((-\alpha,2-\alpha),(2,3-{2\over \alpha});
{y^2\over {\cal
D}})\right) + C_3
\]
in the case $\gamma=0$,
where $y=xt^{-{\alpha\over 2}}$ is the first scale invariant {\rm
(\ref{inv})},
$\phi((\mu,a),(\nu,b);z)$ is the generalized Wright function given by
($\mu+\nu >0$)
\begin{equation}
\label{wright}
\phi((\mu,a),(\nu,b);z):=\sum_{k=0}^\infty {z^k\over \Gamma(a
+\mu k) \Gamma(b+\nu k)},\qquad
\mu,\nu\in \RR,\ a,b\in \CC,
\end{equation}
and $C_1, C_2, C_3$ are arbitrary constants.

\bigskip

For the elements of the theory of the generalized Wright function
(\ref{wright}) including its integral representations and asymptotics
we refer to Wright \cite{Wr2} in the case $\mu,\nu >0$ and to Luchko
and Gorenflo \cite{Luc} in the
case of one of the parameters $\mu,\ \nu$ being negative.

\bigskip

We consider now the equation (\ref{FPDE}) on the semi-axis $x\ge 0$ with
the fractional derivative in the Riemann-Liouville sense:
\begin{equation}
\label{RLFD}
\FTS{\partial^{\alpha}u(x,t)}{\partial t^{\alpha}}=\cases{
\FTS{\partial^{n}u(x,t)}{\partial t^{n}},&$\alpha=n \in\NN$ , \cr
\FTS{1}{\Gamma(n-\alpha)}\FTS{\partial^n}{\partial
t^n}\int\limits^{t}_{0}(t-\tau)^{n-\alpha-1}u(x,\tau)\,d\tau,
&$n-1<\alpha<n$. \cr }
\end{equation}
Also in this case the scale-invariants of  a one parameter group
$T_\lambda$ of scaling transformations for
the equation (\ref{FPDE}) of the form $T_\lambda \circ (x,t,u)=(\lambda
x,\ \lambda^b t,\ \lambda^c u)$ are given by Theorem 3.3.1.

Following Buckwar and Luchko \cite{BucLuc} we restrict ourselves
in the further discussion to the case of the group $T_\lambda$ of
scaling transformations of the form
$T_\lambda \circ (x,t,u)=(\lambda x,\ \lambda^b t,\ u)$. Then the
scale-invariant solutions of the equation (\ref{FPDE}) with the
Riemann-Liouville fractional derivative (\ref{RLFD}) have the form
\begin{equation}
\label{v1}
u(x,t)=v(y),\ y=x t^{-\alpha/2}
\end{equation}
and the substitution (\ref{v1})
reduces the
partial integro-differential equation (\ref{FPDE}) into an ordinary
integro-differential equation with the unknown function $v(y)$ given by
the following theorem.

\theorem{3.3.5}
The reduced equation for the scale-invariant solutions
in the form {\rm (\ref{v1})} of the equation
{\rm (\ref{FPDE})} with the Riemann-Liouville fractional derivative
{\rm (\ref{RLFD})}  is given by
\vskip -6pt
\begin{equation}
\label{red1}
(P_{2/\alpha}^{1-\alpha,\alpha}v)(y)={\cal D}v''(y),\ y>0
\end{equation}
with the left-hand sided Erd\'elyi-Kober fractional differential
operator $P_{\delta}^{\tau,\alpha}$ defined for $0<\delta,\ n-1
<\alpha\le n\in \NN$ by
\vskip -10pt
\begin{equation}
(P_\delta^{\tau,\alpha}g)(y):=\left(\prod_{j=0}^{n -1}
(\tau +j-{1\over \delta} y{d\over dy})\right)
(K_\delta^{\tau+\alpha,n-\alpha} g)(y),\ y>0.
\label{P}
\end{equation}
\vskip -2pt \noindent
Here $(K_\delta^{\tau,\alpha} g)(y)$
is the left-hand sided Erd\'elyi-Kober fractional integral operator {\rm
(\ref{K})}.

\bigskip

The solutions of the equation (\ref{red1}) have been given by
Buckwar and Luchko \cite{BucLuc} for $\alpha \ge 1$.

\theorem{3.3.6}
\label{bucluc}
The scale-invariant solutions of the
equation  {\rm (\ref{FPDE})} with the Riemann-Liouville
fractional derivative {\rm (\ref{RLFD})} in the case $1\le
\alpha<2$ have the form $(y=xt^{-\alpha/2}):$
\vskip -10pt
\begin{equation}
u(x,t)=v(y)\ =\ C_1 \phi(-{\alpha\over 2},1,-y/\sqrt{{\cal D}})+ C_2
\phi(-{\alpha\over 2},1,y/\sqrt{{\cal D}})
\label{sol1}
\end{equation}
with arbitrary constants $C_1,\ C_2$.

\bigskip

\medskip

Now we consider the case $\alpha >2$:

\smallskip

\theorem{3.3.7}
The scale-invariant solutions of the
equation  {\rm (\ref{FPDE})} with the Riemann-Liouville
fractional derivative {\rm (\ref{RLFD})}  in the case $\alpha >2,\
\alpha\not \in{\NN}$ have the form $(y=xt^{-\alpha/2}):$
\begin{equation}
u(x,t)=v(y)=\sum_{j=0}^{[\alpha]}C_j y^{-2+{2\over \alpha}(1+j)}
\mbox{ }_2\Psi_1\left[
 { (1,1),\ (2-{2\over \alpha}(1+j),2) \atop
(\alpha -j,\alpha)}
; {\cal D} y^{-2} \right],
\label{sol3}
\end{equation}
where $C_j,\ 0\le j\le [\alpha]$ are arbitrary constants
and $\mbox{ }_p\Psi_q\left[{(a_1,A_1),\dots,(a_p,A_p)\atop (b_1,B_1)\dots
(b_q,B_q)};z\right]$ is the generalized Wright function {\rm (see \cite{Sri})}:
\begin{equation}
\mbox{ }_p\Psi_q\left[{(a_1,A_1),\dots,(a_p,A_p)\atop (b_1,B_1)\dots
(b_q,B_q)};z\right]=\sum_{k=0}^\infty {\prod_{i=1}^p \Gamma(a_i +A_i k)
\over \prod_{i=1}^q \Gamma(b_i+B_i k)} {z^k\over k!}.
\label{psi}
\end{equation}

\bigskip

In the case  $2<\alpha=n\in {\NN}$
we have the following result.

\bigskip

\theorem{3.3.8}
The scale-invariant solutions of the partial
differential equation $(2<n\in {\NN})$
$$
{\partial^n u\over \partial t^n}
= {\cal D} u_{xx},\ t>0,\ x>0,\ {\cal D}>0
$$
have the form $(y=x/t^{n/2}):$
\begin{equation}
\label{soln}
u(x,t)=\sum_{j=0}^{n-2}C_j y^{-2+{2\over n}(1+j)}\mbox{ }_2\Psi_1\left[
{ (1,1),\ (2-{2\over n}(1+j),2) \atop
(n -j,n)} ; {\cal D} y^{-2} \right]\ +C_{n-1}
\end{equation}
with arbitrary constants $C_j,\ 0\le j \le n-1$.

\bigskip

Finally, following Luchko and Gorenflo \cite{Luc}, we consider
the time-and space-fractional partial differential equation
\vskip -6pt
\begin{equation}
\label{TSF}
\FTS{\partial^{\alpha}u(x,t)}{\partial
t^{\alpha}}={\cal D}\FTS{\partial^{\beta}u(x,t)}{\partial x^{\beta}},
\qquad x > 0, \ t > 0,\ {\cal D} > 0,
\end{equation}
where both fractional derivatives are defined in the
Riemann-Liouville sense (\ref{RLFD}).

\bigskip

\theorem{3.3.9}
The invariants of the group $T_\lambda$ of scaling transformations under
which the equation {\rm (\ref{TSF})} is invariant are given by the
expressions
\vskip -6pt
\begin{equation}
\label{inv_1}
\eta_1(x,t,u)=xt^{-\alpha/\beta},\ \eta_2(x,t,u)=t^{-\gamma}u
\end{equation}
with an arbitrary constant $\gamma$.

\bigskip

\theorem{3.3.10}
The transformation
\vskip -6pt
\begin{equation}
\label{trans}
u(x,t) \ = \  t^\gamma v(y), \quad y \ = \ xt^{-\alpha/\beta}
\end{equation}
reduces the partial differential equation of fractional order
{\rm (\ref{TSF})} to the ordinary differential equation of fractional
order of the form
\vskip -6pt
\begin{equation}
(P_{\beta/\alpha}^{1+\gamma-\alpha,\alpha}
v)(y)={\cal D}y^{-\beta}(D_1^{-\beta,\beta} v)(y),\quad  y>0.
\label{redeq}
\end{equation}
Here the left-hand sided Erd\'elyi-Kober fractional
differential operator $P_\delta^{\tau,\alpha}$ is given by {\rm (\ref{P})}
and the right-hand sided Erd\'elyi-Kober fractional
differential operator $D_\delta^{\tau,\beta}$ is defined
for $0<\delta,\ n-1<\beta\le n\in \NN$ by
\vskip -8pt
\begin{equation}
\label{D}
(D_\delta^{\tau,\beta}g)(y):=\left(\prod_{j=1}^{n}
(\tau +j+{1\over \delta} y{d\over dy})\right)
(I_\delta^{\tau+\beta,n-\beta} g)(y),\ y>0,
\end{equation}
\vskip -2pt \noindent
with the right-hand sided Erd\'elyi-Kober fractional integral operator
\vskip -6pt
\begin{equation}
\label{I_1}
(I_\delta^{\tau,\beta} g)(y):=
\left\{ \begin{array}{ll} {1\over \Gamma(\beta)}
\int_0^1 (1-u)^{\beta -1} u^{\tau}
g(yu^{1/\delta})\, du, & \beta>0,\\ [6pt]
g(y),& \beta=0.
\end{array}
\right.
\end{equation}

\bigskip

Solving the reduced equation we arrive at the following theorem.

\bigskip

\theorem{3.3.11}
Let
\vskip -4pt
$$
{\beta\over 2} \le
\alpha<\beta \le 2,\ \ n-1<\beta\le n \in \NN.
$$
Then the scale-invariant $($according to the transformation
{\rm (\ref{trans})} with $\gamma \ge 0)$ solutions of the partial
differential equation of fractional order {\rm (\ref{TSF})}
have the form
\vskip -8pt
\begin{equation}
\label{solution}
u(x,t)=t^\gamma \sum_{j=1}^n C_j v_j(y), \ y=xt^{-\alpha/\beta},
\end{equation}
\vskip -3pt \noindent
where
\vskip -13pt
\begin{equation}
\label{vj}
v_j(y)=y^{\beta-j}
\phi((-\alpha,1+\gamma-\alpha+{\alpha\over\beta}j),
(\beta,1+\beta-j);y^\beta/{\cal D}),
\end{equation}
the $C_j,\ 1\le j\le n$ are arbitrary real constants, and
$\phi((\mu,a),(\nu,b);z)$ is the generalized Wright function given by
{\rm (\ref{wright})}.

\bigskip

\remark{3.3.3}
In the case $\beta =2$ the scale-invariant solutions
of the equation {\rm (\ref{TSF})} can be expressed in terms of the Wright
function. Indeed, let us consider the linear
combinations of the  solutions {\rm (\ref{vj})}
with $y=xt^{-\alpha/2}$:
$$
u_1(x,t) = t^\gamma(\sqrt{{\cal D}}v_1(y)+v_2(y))
=t^\gamma \phi(-{\alpha\over 2},1+\gamma;y/\sqrt{{\cal D}}),
$$
$$
u_2(x,t) =t^\gamma( -\sqrt{{\cal D}}v_1(y)+v_2(y))
=t^\gamma \phi(-{\alpha\over 2},1+\gamma;-y/\sqrt{{\cal D}}).
$$
These scale-invariant solutions are given in Theorem 3.3.6
in the case $\gamma =0$.

\bigskip

\remark{3.3.4}
For $0<\beta\le 1$ the equation {\rm (\ref{TSF})} has only one
solution which is scale-invariant with respect to the
transformation {\rm (\ref{trans})}. This solution has the form
$(y=xt^{-\alpha/\beta}):$
\vskip -10pt
$$
u(x,t)=t^\gamma v_1(y) = t^\gamma y^{\beta-1}\,
\phi((-\alpha,1+\gamma-\alpha+{\alpha\over\beta}),(\beta,\beta);
y^\beta/{\cal D}).
$$
In the case $\beta =1$, this function is expressed in terms of
the Wright function ($y=xt^{-\alpha}$):
\vskip -10pt
$$
u(x,t)=t^\gamma v_1(y)=t^\gamma \phi(-\alpha,1+\gamma;y/{\cal D}).
$$

\section*{Acknowldgements}
The Authors acknowledge  partial support by the the Research
Commission of Free University of Berlin
(Project "Convolutions") and by the Italian CNR and INFN.
The paper was presented at the 3rd Workshop TMSF
(Transform Methods and Special Functions), Sofia, Bulgaria, 1999.

\end{document}